\documentclass[a4paper,11pt]{article}
\pdfoutput=1 

\usepackage{jheparxiv} 

\usepackage[T1]{fontenc} 

\usepackage{tensor}
\usepackage{amsmath}
\usepackage{amssymb}
\usepackage{mathrsfs}
\usepackage{graphicx}
\usepackage{stmaryrd}
\usepackage{twistor}
\usepackage{commath}

\newcommand{\veps}{\varepsilon}

\newcommand{\msf}[1]{\mathsf{#1}}

\newcommand{\mA}{\mathcal{A}}

\title{The 4d/2d correspondence in twistor space and holomorphic Wilson lines}

\author[a]{Wei Bu}\emailAdd{w.bu@sms.ed.ac.uk}
\author[b]{and Eduardo Casali}\emailAdd{ecasali@g.harvard.edu}
\affiliation[{a}]{School of Mathematics and Maxwell Institute for Mathematical Sciences\\
University of Edinburgh, EH9 3FD, UK}
\affiliation[{b}]{Center for the Fundamental Laws of Nature, Harvard University,\\Cambridge, MA 02138, USA}

\abstract{We give an explicit realization of the 4d local operator /  2d conformal block correspondence of Costello and Paquette in the case of gauge theories. This is accomplished by lifting the 4d local operators to non-local operators in twistor space using a holomorphic generalization of the Wilson line. This procedure automatically constructs the 2d conformal blocks corresponding to the local operator. We interpret this lifting as effectively integrating out the 2d degrees of freedom living on the defect. We present some 2d chiral CFT representation of the defect algebra whose correlators reproduce the conformal blocks obtained by the lifting procedure.}

\begin{document} 
\maketitle
\flushbottom
\section{Introduction}

The search for a holographic correspondence in asymptotically flat spaces has found renewed interest spurred by results in gravitational amplitudes tying together asymptotic symmetries and soft factors \cite{He:2014laa,Strominger:2013jfa,He:2014cra,He:2015zea,Kapec:2014opa,Kapec:2016jld,Pasterski:2016qvg,Campiglia:2015kxa,Strominger:2013lka,Campiglia:2015qka,Pasterski:2017kqt,Strominger:2014pwa,Ball:2019atb,He:2019jjk,Henneaux:2019yqq}. This line of inquiry goes by the name of celestial holography, positing the existence of a 2d theory living on the celestial sphere at asymptotically null infinity dual to a gravitational theory in the 4d bulk. An even larger algebra of generalized soft currents was found in~\cite{Guevara:2021abz}, inspired by rewriting the S-matrix in a basis adapted to the asymptotic symmetries~\cite{Pasterski:2017kqt}. This larger soft algebra suffers from ambiguities, rendering it well defined when either only positive, or only negative helicity particles are present. One half of this algebra, for example the positive helicity half, was identified as the loop algebra of the wedge algebra of $w_{1+\infty}$ \cite{Strominger:2021mtt}. Interestingly, the same algebra appears in the nonlinear graviton construction of Penrose for self-dual spacetimes \cite{Penrose:1967wn} as the diffeomorphism algebra of a 2-plane preserving a Poisson structure. Using a twistor space representation for the scattering amplitudes, the authors of~\cite{Adamo:2021lrv} showed that this larger soft-symmetry and the algebra of Poisson-diffeomorphisms are the same. Further works showed how twistor strings can be used to compute these soft algebras from the worldsheet CFT \cite{Adamo_2022,Bu:2021avc} where the soft-expansion of the vertex operators is explicitly given by generators of the Poisson-diffeomorphism algebra. A similar story is also present in Yang-Mills, where a tower of soft symmetries can be shown to be equivalent to the Ward construction \cite{Ward:1977ta} in twistor space for self-dual Yang-Mills solutions. It is well established by now that twistor methods can be of great usefulness in the study of amplitudes and there is an ample literature on twistor actions and twistor strings \cite{Mason:2005zm,Boels:2006ir,Mason:2007ct,Witten:2003nn,Berkovits:2004jj,Skinner:2013xp}. It is then not surprising that some results from celestial holography can be connect to constructions in twistor space. 

A very interesting application appeared recently in~\cite{Costello:2022wso} where the authors considered holomorphic theories in twistor space. Using twisted holography \cite{Costello:2018zrm,Costello:2020jbh}, a toy model of the AdS/CFT holography, they showed how the the algebra of currents living on a 2d defect is the same as the $w_{1+\infty}$ algebra of the soft sector of 4d scattering amplitudes. Moreover, in their construction the negative helicity currents can be coupled consistently, though now commuting among themselves. Of main interest for our work is another related result of that work: there is a correspondence between local operators in a 4d theory and conformal blocks for the 2d defect algebra. This correspondence is mediated through the 6D theory in twistor space which is used in the proofs presented by the authors. These proofs are non-constructive and the examples of the correspondence given in \cite{Costello:2022wso} rely on some guess work and knowledge of the perturbative expansion of the 4d theory. To better understand this correspondence an explicit procedure on how to construct the conformal blocks given a local operator would be very useful. The main result of this paper is such a  construction in the case of gauge theories. 

In this work we show how the correspondence between local operators and conformal blocks in the case of Yang-Mills theories can be seen as a simple consequence of lifting 4d local operators to operators in the 6D theory living on a linearly embedded $\mathbb{CP}^1$. Using some less well-known tools from the amplitudes literature we give an explicit procedure that generates the conformal blocks corresponding to a local operator. We also show how the procedure has to be modified for theories with supersymmetry, exemplified by the $\mathcal{N}=4$ case which we consider in some detail. Consequently, we learn how the framework introduced in \cite{Costello:2022wso} gives a novel interpretation to some twistor constructions in the amplitudes literature as originating from integrating out the 2d degrees of freedom. We give some examples of this using chiral CFT representations for the degrees of freedom on the defect. In the case of $\mathcal{N}=4$ SYM this turns out to be identical to a sector of the original twistor string \cite{Witten:2003nn,Berkovits:2004jj}.

This paper is organized as follows: Section \ref{action_review} starts with a review on twistor space and twistor actions for gauge theories and ends with a review on the 4d/2d correspondence of \cite{Costello:2022wso}. In section \ref{lifting_section} we describe how to lift local operators in 4d to twistor space. This construction is used in section \ref{blocks} to obtain explicitly the conformal blocks corresponding to local operators. We finish in section \ref{discussion} with some words on generalizations and some open questions.

\section{Twistor space and twistor actions}\label{action_review}

We represent twistor space as an open subset of the complex projective 3-space $\mathbb{CP}^3$ using $Z^I\in\mathbb{C}^4$ to define homogeneous coordinates on $\mathbb{CP}^{3}$ by the relation $Z^I\sim s\;Z^I$, with $s\in\mathbb{C}^*$ a complex scaling. We divide them into a pair $Z^I=(\mu^{\dot\alpha},\lambda_{\alpha})$ with $\dot\alpha,\;\alpha$ spin $\frac{1}{2}\times\frac{1}{2}$ representations of $SL(2|\mathbb{C})\times SL(2|\mathbb{C})$ considered as the complexification of the Lorentz algebra. Twistor space $\mathbb{PT}$ is obtained by removing the complex line $\lambda=0$ from $\mathbb{CP}^3$, leaving an open subset which can be identified with the total space of a $\mathcal{O}(1)\oplus\mathcal{O}(1)$ fibration over $\mathbb{CP}^1$. This is easily seen by defining non-homogeneous coordinates in a patch on $\mathbb{CP}^1$ by $z=\lambda_{1}/\lambda_{0}$ and on the fibers $v^{\dot\alpha}=\mu^{\dot\alpha}/\lambda_{0}$. We will mostly use homogeneous coordinates but might freely change to the non-homogeneous coordinates depending on the context.

The relation between twistor space and spacetime is non-local and given by the incidence relation
\begin{equation}\label{incidence}
    \mu^{\dot\alpha}=x^{\dot\alpha\alpha}\lambda_{\alpha}.
\end{equation}
This realizes (complexified) spacetime as the moduli space of degree one (complex) lines in twistor space.  Real structures on spacetime can be given by choosing appropriate complex conjugations of $Z$, different conjugations giving rise to different signatures for the 4d metric \cite{Woodhouse:1985id}. The relation is particularly simple in Euclidean signature, where $\mathbb{PT}$ can be seen as a sphere bundle over spacetime $\mathbb{R}^4\times\mathbb{CP}^1$. Although we mostly work in the complex category, the Euclidean construction is helpful to have in mind for our purposes. 

Fields parameterised by homogeneous coordinates descend straightforwardly to twistor space if they have zero net weight on $Z$, that is, $a(s\,Z)=a(Z)$ is a well defined field on $\mathbb{PT}$. Fields with defined scaling weight $g(s\,Z)=s^{n}g(Z)$ descend to the same class of fields on twistor space but twisted by $\mathcal{O}(n)$. A theorem by Penrose~\cite{Penrose:1967wn} relates the cohomology classes $H^{1}(\mathbb{PT},\mathcal{O}(-2+2n))$ in twistor space to helicity $n$ solutions of the free massless equation in spacetime. Dolbeault representatives for the cohomology classes $H^1(\mathbb{PT},\mathcal{O}(n))$ are given by $\bar\partial$-closed antiholomorphic one-forms up to the addition of $\bar\partial$-exact forms. These can be use to write down actions in twistor space, which usually take the form of holomorphic BF systems. For example we can pair a $(0,1)$-form field $a\in\Omega^{0,1}(\mathbb{PT})$ of zero weight with another $(0,1)$-form field $g\in\Omega^{0,1}(\mathbb{PT},\mathcal{O}(-4))$ with the action
\begin{equation}
    S_{Maxwell}=\int_{\mathbb{PT}}D^{3}Z\wedge g\wedge\bar\partial a
\end{equation}
where $D^3Z=\varepsilon_{IJKL}Z^IZ^JZ^kZ^L$ denotes the top holomorphic form on $\mathbb{PT}$. This form is twisted by $\mathcal{O}(4)$ so the whole integrand above is weightless and well defined over twistor space. This action is invariant under the gauge symmetries
\begin{equation}\label{ag_eom}
    a\rightarrow a+\bar\partial c_0\quad g\rightarrow g+\bar\partial c_{-4}
\end{equation}
with $c_0$ a function in twistor space and $c_{-4}$ a section of $\mathcal{O}(-4)$. The equations of motion $\bar\partial a=0=\bar\partial g$ together with the gauge redundancy \eqref{ag_eom} imply that $a\in H^1(\mathbb{PT})$ and $g\in H^1(\mathbb{PT},\mathcal{O}(-4))$. Hence, $a$ describes a positive helicity photon while $g$ describes a negative helicity photon.

The generalization to an interacting Yang-Mills theory is straightforward. The field $a$ is promoted to a partial connection on a gauge bundle $E$ over $\mathbb{PT}$ for some gauge group $G$, which we can take to be $SU(N)$ for simplicity. The field $g$ is taken to have values in the adjoint representation of $G$, and partial derivatives are promoted to covariant ones. The resulting action is
\begin{equation}\label{action_SDYM}
    S_{SDYM}=\int_{\mathbb{PT}}D^3Z\wedge \tr\left(g\wedge \bar\partial a+g\wedge a\wedge a\right)=\int_{\mathbb{PT}}D^3Z\wedge \tr \left(g\wedge F(a)\right)
\end{equation}
where $F(a)=\bar\partial a +a\wedge a $ is the (0,2) component of the field strength. 
The on-shell equations impose the vanishing of the antiholomorphic component of the curvature giving a holomorphic bundle over twistor space. A theorem by Ward relates holomorphic bundles over $\mathbb{PT}$ to self-dual solutions to the Yang-Mills equations on spacetime \cite{Ward:1977ta}. The field $g$ can be considered as a perturbation of the opposite helicity over this non-linear self-dual background.

The action~\eqref{action_SDYM} suffers from a quantum anomaly in twistor space related to the all-plus one-loop amplitude. A simple way to see this is from its connection with the Chalmers \& Siegel action~\cite{Chalmers:1996rq}, which is a spacetime action for self dual Yang-Mills
\begin{equation}\label{CS_action}
S_{CS}=\int_{\mathbb{R}^4}\d^4x\,\tr\left( B_{\alpha\beta}F^{\alpha\beta}(A)\right)=\int_{\mathbb{R}^4}\d^4x\,\tr\left(B_{\alpha\beta}(\partial^{\dot\alpha(\alpha}A^{\beta)}{}_{\dot\alpha}+[A^{\dot\alpha(\alpha},A^{\beta)}{}_{\dot\alpha}])\right),
\end{equation}
where $(\alpha\beta)$ indicates symmetrization of the spinor indices. The field $B_{\alpha\beta}$ is an anti-self-dual two form and $F^{\alpha\beta}$ is the anti-self-dual part of the curvature for a gauge field on spacetime. The equations of motion set the anti-self-dual part of the curvature to zero giving a purely self-dual connection. The twistor action~\eqref{action_SDYM} was shown to be classically equivalent to the Chalmers \& Siegel action by integrating over the $S^2$ fiber~\cite{Boels:2006ir,Mason:2005zm}. If this equivalence holds also at the quantum level the amplitudes computed with either theory must coincide. The Chalmers \& Siegel action has a non-zero one-loop all-plus amplitude which can be seen to be absent from the twistor string in an axial gauge $Z_*^Ia_I=0$, where the action is free. This anomaly has an origin in the chiral nature of the theory on twistor space and can be cured in several different ways~\cite{Costello:2021bah,Costello:2022wso}. In particular, Yang-Mills theories with any amount of supersymmetry have vanishing all-plus amplitudes and so are anomaly free in twistor space.

In this work we will only consider twistor actions for gauge theories, but we note that that there are also proposals for conformal and Einstein gravity \cite{Adamo:2013cra,Sharma:2021pkl,Adamo:2013tja,Mason:2005zm} for which a generalization of the methods employed here should be possible. 

\subsection{The 4d local operator / 2d conformal block correspondence}

The authors in \cite{Costello:2022wso} showed that theories in twistor space give rise to a 4d/2d correspondence. On the 4d side are local operators of a theory obtained by dimensional reduction of the twistor action on the $S^2$ fiber. On the 2d side are conformal blocks for chiral algebras that live on holomorphic line defects in twistor space. We will give an abbreviated version of the arguments presented in \cite{Costello:2022wso}, referring the reader to that work for more details.

The argument goes as follows: Consider a real codimension one surface in twistor space $S^3\times S^2\subset\mathbb{R}^4\times S^2\simeq\mathbb{PT}$. This is the quantization surface where the Hilbert space $\mathcal{H}(S^3\times S^2)$ of the theory on twistor space lives. Dimensionally reducing along the $S^2$ fiber preserves all the degrees of freedom\footnote {The $S^2$ is essentially an auxiliary direction, see \cite{Boels:2006ir}} therefore $\mathcal{H}(S^3\times S^2)=\mathcal{H}(S^3)$. The latter is the Hilbert space of the 4d theory built out of the action of local operators on the vacuum vector. Going back to twistor space, dimensionally reduce instead on $S^3$ and keep all the KK modes. In this way $\mathcal{H}(S^3\times S^2)=\mathcal{H}^{KK}(S^2)$ which is a Hilbert space for a 3 real dimensional theory containing all of the $S^3$ harmonics. This gives the equality $\mathcal{H}(S^3)=\mathcal{H}^{KK}(S^2)$ between Hilbert spaces and what is left to do is to characterize the Hilbert space on $S^2$.

In the case of self-dual Yang-Mills and self-dual gravity on twistor space the Hilbert space $\mathcal{H}^{KK}(S^2)$ was identified in \cite{Costello:2022wso} by explicitly performing the dimensional reduction on $S^3$ of the the respective twistor actions. The self-dual Yang-Mills action \eqref{action_SDYM}\footnote{Their action also contained a coupling to a scalar field in order to cancel the one loop anomaly. We instead prefer to work with a supersymmetric theory.} reduces to a three real dimensional action of a topological-holomorphic theory of the form
\begin{equation}
\int_{\mathbb{R}^*\times\mathbb{CP}^1}d z\wedge\tr\left( B\wedge F(A)+\eta D_A\phi\right)+\text{KK modes}.
\end{equation}
where the topological direction is $\mathbb{R}^*$ direction~\cite{Aganagic:2017tvx} and the holomorphic directions are along  $S^2$. In the presence of a two dimensional boundary, these theories develop a chiral algebra along the boundary. Moreover the Hilbert space of the 3d theory can be identified with the space of conformal blocks for this chiral algebra, see \cite{Aganagic:2017tvx,Costello:2020jbh} for details. A similar construction holds for self-dual gravity~\cite{Costello:2022wso}.

The chiral algebra for self-dual Yang-Mills is generated by two adjoint-valued fields: $J^\msf{a}$ which couples to the bulk gauge field $a$, and $\tilde{J}^\msf{a}$ coupling to the bulk field $g$. Their chiral algebra is given by the OPEs
\begin{align}
    &J^\msf{a}(z_1)J^\msf{b}(z_2)\sim \frac{f^{\msf{a}\msf{b}}{}_\msf{c}}{z_1-z_2}J^\msf{c}(z_2) \label{ca_sdYM} \\
    &J^\msf{a}(z_1)\tilde{J}^\msf{b}(z_2)\sim \frac{f^{\msf{a}\msf{b}}{}_\msf{c}}{z_1-z_2}\tilde{J}^\msf{c}(z_2) \label{ca_+-}
\end{align}
and a regular $\widetilde{J}\widetilde{J}\sim0$ OPE. In this case, the 2d/4d correspondence is between the local operators of self-dual Yang-Mils, generated by polynomials on the fields $B^{\alpha\beta}$, $F_{\dot\alpha\dot\beta}$ in~\eqref{CS_action} and their derivatives modulo equations of motion, and conformal blocks for the chiral algebra~\eqref{ca_sdYM}. 

As an example, take the local operator $\tr B^2$ on spacetime. It can be added to the action \eqref{CS_action} to recover the full Yang-Mills theory and so can be used as an interaction vertex to compute observables perturbatively around the self-dual sector. According to the 2d/4d correspondence, this operator corresponds to a conformal block $\langle\text{tr}B^2|$ which was argued in~\cite{Costello:2022wso} to act on elements of the algebra~\eqref{ca_sdYM} as
\begin{equation}
\langle\text{tr}B^2|\tilde{J}^{\msf{a}_1}(z_1)\tilde{J}^{\msf{a}_2}(z_2)J^{\msf{a}_3}(z_3)\cdots J^{\msf{a}_n}(z_n)\rangle=
\tr(T^{\msf{a}_1}T^{\msf{a}_2}\cdots T^{\msf{a}_n})\,\frac{z^4_{12}}{z_{12}z_{23}\cdots z_{n1}} +\dots
\end{equation}
where $z_{ij}=z_i-z_j$, the $T^{\msf{a}}$s are the generators of the gauge group, and we only wrote one ordering explicitly. Here we identifying the holomorphic coordinates $z_i$ with the chiral spinors $\lambda_i$\footnote{This identification can be made explicit by using twistor wavefunctions in the plane wave basis.}, which is proportional to $k_i$ for massless momenta $p_{\alpha\dot\alpha}=k_{\alpha}\tilde{k}_{\dal}$ when evaluated on momentum eigenstate. Then this correlator is the same as the prefactor of MHV amplitudes \cite{Nair:2005iv}. These are usually computed by summing many Feynman diagrams with one insertion of $\text{tr} B^2$ and the three-point interaction from \eqref{CS_action}. The fact that the conformal block reproduces the prefactor of this amplitude is no coincidence and follows from a construction on twistor space which will be described in the next section. 

There are other operators in the chiral algebra which are dual to the KK modes. These sit in representations of the $SL(2|\mathbb{C})$ acting on dotted indices since the KK modes come from polynomials in $\mu^{\dot\alpha}$. In \cite{Costello:2022wso} these are labelled $J[m,n]$ for weight $(m-n)/2$ and spin $(m-n)/2$ representations corresponding to the KK modes of the positive helicity gluon. Similarly, $\tilde{J}[m,n]$ are the chiral algebra duals for the modes of the negative helicity gluon. Their OPEs are
\begin{align}
&J^\msf{a}[m,n](z_1)J^\msf{b}[k,l](z_2)\sim\frac{f^{\msf{a}\msf{b}}{}_\msf{c}}{z_1-z_2}J^\msf{c}[m+k,n+l](z_2)\\
&J^\msf{a}[m,n](z_1)\tilde{J}^\msf{b}[k,l](z_2)\sim\frac{f^{\msf{a}\msf{b}}{}_\msf{c}}{z_1-z_2}\tilde{J}^\msf{c}[m+k,n+l](z_2)
\end{align}
The algebra of the $J[m,n]$ currents has recently been shown to arise from the soft sector of Yang-Mills amplitudes~\cite{Strominger:2021mtt,Guevara:2021abz} and has been argued to be a symmetry of a putative holographic dual to it in flat space. This is also the symmetry algebra of the self-dual sector of Yang-Mills and has a natural geometric interpretation in twistor space~\cite{Adamo_2022,Bu:2021avc,Adamo:2021lrv}.

Chiral algebras dual to gravitational fields in the self-dual sector where also discussed in~\cite{Costello:2022wso} where a similar local operator / conformal block correspondence holds. The explicit realisation of this duality in this case seems more subtle than for gauge theories and will be discussed in future work. Other gauge theories should also have associated chiral algebras which can be found by applying the same reasoning as above to the twistor actions in \cite{Mason:2007ct}. We exemplify this in section \ref{N=4SYM} with the maximally supersymmetric Yang-Mills theory which has a natural description in twistor space.

\section{Lifting local operators to twistor space}\label{lifting_section}

Generic local operators in spacetime don't normally lift to local operators in twistor space. The relationship between these spaces is non-local \eqref{incidence} and we expect the description of local operators living at a point $x$ in spacetime to be given by some degrees of freedom living on the holomorphic \textit{line} $L_x\cong\mathbb{CP}^1$ in twistor space corresponding to the spacetime point $x\in\mathbb{R}^4$. 


To exemplify the construction we will first consider local operators in self-dual Yang-Mills theory. In spacetime its action is given by the action~\eqref{CS_action} of Chalmers \& Siegel~\cite{Chalmers:1996rq} while in twistor space it is given by the holomorphic BF action \eqref{action_SDYM}. 
 We're looking to lift the local operator $\text{tr} B^2(x)$ to twistor space. The field $B$ corresponds to a negative helicity gluon and is described in twistor space by the field $g^\msf{a}\in\Omega^{(0,1)}(\mathbb{PT},\text{End}(E)\times\mathcal{O}(-4))$, where $E$ is a gauge bundle fibred over twistor space. At the linear level\footnote{The non-linear version of this correspondence is given by Ward \cite{Ward:1977ta} and requires the construction of Holomorphic frames in twistor space.}, this is given by the Penrose transform
 \begin{equation}\label{PT_g}
 B^\msf{a}_{\alpha\beta}(x)=\int_{\mathbb{CP}^1}\la\lambda\d\lambda\ra\lambda_{\alpha}\lambda_{\beta}\,g^{\msf{a}}(Z)|_{L_x} =\int_{\mathbb{CP}^1}\langle\lambda\d\lambda\rangle\lambda_\alpha\lambda_\beta\, g^\msf{a}(x^{\alpha\dot\alpha}\lambda_\alpha,\lambda)
 \end{equation}
where $\la\lambda\d\lambda\ra=\varepsilon^{\gamma\delta} \lambda_{\gamma}d\lambda_{\delta}$, and the powers of $\lambda$ make up the correct weight such that the integrand is weightless and well-defined on $\mathbb{CP}^1$. Both operators in $\text{tr} B^2$ are inserted at the same spacetime point $x$, hence by the incidence relation, they sit on the same line $L_x$ in twistor space. However there is no reason for them to sit at the same point in $L_x$. We represent them with fields, $g^\msf{a}(Z_i)=g^\msf{a}(\mu_i,\lambda_i)=g^\msf{a}(x\lambda_i,\lambda_i)$ with $i=1,2$, inserted at two different points on the line. The dependence on these points is arbitrary and the end result should not depend on them. Moreover, we also need to specify how to compare the gauge bundle at different points in $L_x$ to be able to multiply the operators and take their trace.

\subsection{The holomorphic Wilson line}

This issue was tackled in the literature on scattering amplitudes in twistor space \cite{Mason:2010yk,Bullimore:2011ni} in the study of the relation between holomorphic linking and scattering amplitudes. For operators in a gauge theory in spacetime sitting at distinct points, a gauge invariant combination can be given by connection them through a Wilson line. In twistor space we only have access to a partial connection given by the positive helicity field $a^{\msf{a}}\in\Omega^{0,1}(\mathbb{PT},\text{End}(E))$. Lacking a full connection, the next best thing we can do is to define a \textit{holomorphic} analogue of a Wilson line. We will describe the construction by analogy with the usual Wilson loop, further details can be found in the aforementioned works.

Given a connection $A$, and a path $\gamma:[0,1]\rightarrow \mathbb{R}^n$ between two points $y_1=\gamma(0)$ and $y_2=\gamma(1)$, the usual Wilson line is given by the path ordered exponential
\begin{equation}
    W_{\gamma}[y_1,y_2]=\text{Pexp}\left(-\int_\gamma A\right)
\end{equation}
where the connection $A$ has been pulled back to the path $\gamma$. Its definition is given by the series
\begin{multline}
W_{\gamma}[y_1,y_2] = 1 - \int_{y_1}^{y_2}\d y A(y)+\int_{y_1}^{y_2}\d y \int_{y_1}^{y}\d y' A(y)A(y')+ \cdots\\
=1+\sum_{n=1}^\infty(-1)^n\int_{[0,1]^n}\d s_n\cdots\d s_1 A(s_n)\frac{\d \gamma}{\d s_n}\theta(s_n-s_{n-1})\cdots\theta(s_2-s_1)A(s_1)\frac{\d\gamma}{\d s_1}
\end{multline}
with a kernel obeying 
\begin{equation}
\frac{\d}{d s}\theta(s-s')=\delta(s-s').
\end{equation}

In the holomorphic case we only have the $(0,1)$ part of the connection, and it is not at first clear what a holomorphic path between two points is. The solution is to consider the parallel transport over a holomorphically embedded line as the generalization of the path $\gamma$ using the appropriate kernel. The kernel should be a (1,0) form so that we can pair it with the partial connection and integrate over $\mathbb{CP}^2$. Given two points $z_1$ and $z_2$ on a sphere $\mathbb{CP}^1$ we define the meromorphic differential
\begin{equation}\label{mero_dif}
    \omega_{z_1,z_2}=\frac{1}{2\pi\im}\,\frac{(z_1-z_2)\d z}{(z_1-z)(z-z_2)}
\end{equation}
with residue $1$ at $z_1$ and $-1$ at $z_2$. It obeys
\begin{equation}\label{inverse}
\bar\partial\omega_{z_1,z_2}=\delta^2(z-z_1)-\delta^2(z-z_2)
\end{equation}
and gives a natural holomorphic generalization of the real kernel. Given an abelian partial connection $a$, we can use this to define an abelian holomorphic Wilson line by
\begin{equation}
    W[z_1,z_2]=\text{exp}\left(-\int_{\mathbb{CP}^1}\omega_{z_1,z_2}\wedge a\right)
\end{equation}
This gives the parallel transport of an holomorphic bundle from $z_1$ to $z_2$. The non-abelian holomorphic Wilson line the follows straightforwardly by defining a holomorphically path ordered exponential
\begin{multline}\label{Wilson_line}
W[z_1,z_2]=\,\text{P}\text{exp}\left(-\int_{\mathbb{CP}^1}\omega_{z_1,z_2}\wedge a\right)
=1+\sum_{m=1}^{\infty}(-1)^m\int_{(\mathbb{CP}^1)^m}\bigwedge_{i=1}^m\omega_i \wedge a_i\,(\sigma_i)
\end{multline}
where the last term was written in an abbreviated from. It is given by
\begin{multline}\label{Wilson_line_def}
    \bigwedge_{i=1}^m\omega_i \wedge a_i(\sigma_i) =\omega_{z_2,\sigma_m}\wedge\omega_{\sigma_m,\sigma_{m-1}}\wedge\cdots\wedge\omega_{\sigma_1,z_1}\wedge a_1(\sigma_1)\wedge\dots\wedge a_m(\sigma_m) \\
    =\frac{(z_2-z_1)}{(z_2-\sigma_m)(\sigma_m-\sigma_{m-1})\cdots(\sigma_1-z_1)}\bigwedge_{i=1}^m a_i(\sigma_i)\d\sigma_i.
\end{multline}
This Wilson line obeys the same properties as expected of their real category counterpart, providing the parallel transport of the partial connection and transforming in the adjoint of the gauge group at its endpoints. When considering the sphere as a holomorphically embedded line in twistor space the Wilson line will also depend on the embedding map $\mathcal{C}:\mathbb{CP}^1\rightarrow\mathbb{PT}$ through the pullback of the gauge connection $a$. For our applications we only need to consider linearly embedded lines and will identify the coordinates of the sphere with its image in twistor space.


\subsection{Lifting operators}

With the holomorphic Wilson line in hand we can now write down the twistor space lift of local spacetime operators. It is worth noting that besides the holomorphic Wilson line, there has been other descriptions of spacetime local operators on twistor space \cite{Chicherin:2014uca,Chicherin:2016soh,Adamo:2011cd,Adamo:2011dq,Koster:2016fna}. Given two fields $g(x\lambda_1,\lambda_1)$ and $g(x\lambda_2,\lambda_2)$ on the same $\mathbb{CP}^1$ we can connect them together using the holomorphic Wilson lines. In this way the lift of $\text{tr} B^2$ to twistor space looks like
\begin{equation}
   \text{tr}(B^2)\rightarrow \text{tr}\,\left( g(x\lambda_1,\lambda_1)W[\lambda_1,\lambda_2]g(x\lambda_2,\lambda_2)W[\lambda_2,\lambda_1]\right)
\end{equation}
where we reverted to homogeneous coordinates. The last thing to do is to integrate over the two points on the sphere since the spacetime operators don't depended on these points
\begin{equation}\label{trb2}
    \text{tr} B^2\rightarrow\int_{(\mathbb{CP}^1)^2}\langle\lambda_1\d\lambda_1\rangle\langle\lambda_2\d\lambda_2\rangle\langle\lambda_1\lambda_2\rangle^2\,\text{tr}\,\left( g(x\lambda_1,\lambda_1)W[\lambda_1,\lambda_2]g(x\lambda_2,\lambda_2)W[\lambda_2,\lambda_1]\right)
\end{equation}
where $\langle\lambda_1\lambda_2\rangle^2$ was restored from the Penrose transform~\eqref{PT_g}, and so, cancels the leftover projective weights of the $g$ fields. 

\begin{figure}[h]
    \centering
    \includegraphics[width=0.8\textwidth]{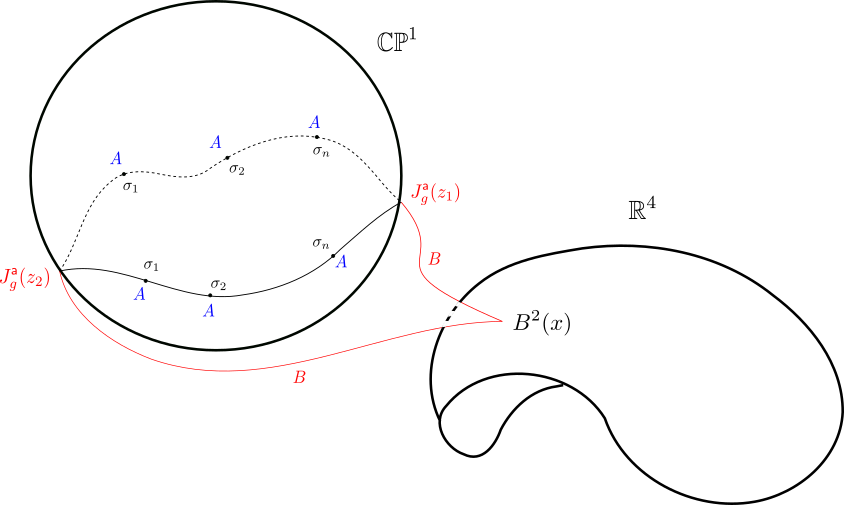}
    \caption{Lift of a local operator to twistor space using the holomorphic Wilson line.}
    \label{g}
\end{figure}

To show the equivalence explicitly we use a partial gauge fixing available in twistor space~\cite{Woodhouse:1985id} in which the component of the connection $a$ along the $S^2$ direction in the fibration $\mathbb{R}^4\times S^2$ is trivial. In this gauge the expression \eqref{trb2} reduces to
\begin{multline}
\int_{(\mathbb{CP}^1)^2}\langle\lambda_1\d\lambda_1\rangle\langle\lambda_2\d\lambda_2\rangle\langle\lambda_1\lambda_2\rangle^2\,\text{tr}\, g(x\lambda_1,\lambda_1)g(x\lambda_2,\lambda_2)\\=\epsilon^{\alpha\beta}\epsilon^{\gamma\delta}\tr\int_{\mathbb{CP}^1}\langle\lambda_1\d\lambda_1\rangle \lambda_{1,\alpha}\lambda_{1,\gamma}\,g(x\lambda_1,\lambda_1)\int_{\mathbb{CP}^1}\langle\lambda_2\d\lambda_2\rangle \lambda_{2,\beta}\lambda_{2,\delta}\,g(x\lambda_2,\lambda_2)=\text{tr}B^2
\end{multline}
where we used the Penrose transform in the second line. The generalization to any polynomial on $B$ follows in a straightforward way: the twistor fields $g$ are linked by the holomorphic Wilson lines and the contractions of the spacetime indices dictate the powers of $\langle\lambda_i\lambda_j\rangle$ that makes the integrand weightless. For example:
\begin{equation}
\text{tr} B_{\alpha\beta}B_{\gamma\delta}B^{\alpha\beta}B^{\gamma\delta}\rightarrow \langle13\rangle^2\langle24\rangle^2\text{tr } g_1W_{12}g_2W_{23}g_3W_{34}g_4W_{41}
\end{equation}
while
\begin{equation}
\text{tr} B^\alpha{}_\beta B^\beta{}_\gamma B^\gamma{}_\delta B^\delta{}_\alpha\rightarrow \langle12\rangle\langle23\rangle\langle34\rangle\langle41\rangle\text{tr } g_1W_{12}g_2W_{23}g_3W_{34}g_4W_{41}.
\end{equation}
Operators containing derivatives of $B$ are described in twistor space by trading the spacetime derivatives with $\mu$ derivatives using the incidence relation. At the linear level, the correspondence is\footnote{The non-linear correspondence makes use of holomorphic frames which are trivial in this partial gauge-fixing.}
\begin{equation}
\frac{\partial}{\partial x^{\alpha\dot\alpha}}B_{\beta\gamma}=\int \langle\lambda\d\lambda\rangle \lambda_\alpha\lambda_\beta\lambda_\gamma\frac{\partial}{\partial\mu^{\dot\alpha}}g(\mu,\lambda)|_{\mu^{\dot\alpha}=x^{\alpha\dot\alpha}\lambda_\alpha}
\end{equation}
and the operator $\frac{\partial}{\partial\mu^{\dot\alpha}}g=\partial_{\dot\alpha}g$ can be connected using the Wilson line as before since the gauge transformations on the line depend only on $\lambda$. For example, 
\begin{equation}
\tr\,\partial_{\alpha\dot\alpha}B_{\gamma\delta}\partial^{\delta\dot\alpha}B^{\gamma}{}_{\sigma}B^{\sigma\alpha}\rightarrow\langle12\rangle^2\langle13\rangle\langle23\rangle \tr\,\partial_{\dot\alpha}g_1W_{12}\partial^{\dot\alpha}g_2W_{23}g_3W_{31}
\end{equation}


Similarly, we can promote local operators containing the curvature $F_{\dot\alpha\dot\beta}(x)=\partial_{\alpha(\dot\alpha}A_{\dot\beta)}{}^{\alpha}+A_{\alpha(\dot\alpha}A_{\dot\beta)}{}^\alpha$ to twistor space by using the Penrose transform
\begin{equation}
\frac{\partial}{\partial x^{\alpha(\dot\alpha}}A_{\dot\beta)}{}^\alpha=\int_{\mathbb{CP}^1}\langle\lambda\d\lambda\rangle\frac{\partial}{\partial\mu^{\dot\alpha}}\frac{\partial}{\partial\mu^{\dot\beta}}a(x\lambda,\lambda)
\end{equation}
for their linearized part, the expansion of the Wilson line provides the non-linear pieces\footnote{Gauge invariance is more subtle in this case, but the prescription gives gauge invariant form-factors as described in \cite{Koster:2016ebi,Koster:2016loo}.}. For example, $\tr F^2$ is represented by two $F$s at different position in the $\mathbb{CP}^1$ fibre connected with holomorphic Wilson lines
\begin{equation}
    \text{tr} F^2\rightarrow\int_{(\mathbb{CP}^1)^2} \langle\lambda_1\d\lambda_1\rangle\langle\lambda_2\d\lambda_2\rangle \text{tr} \left( \frac{\partial^2}{\partial\mu^{\dal}\partial\mu^{\dot\beta}} a(x\lambda_2,\lambda_2)W[\lambda_2,\lambda_1]\frac{\partial^2}{\partial\mu_{\dal}\partial\mu_{\dot\beta}}a(x\lambda_1,\lambda_1)W[\lambda_1,\lambda_2] \right).
\end{equation}
Derivatives of $F$ are treated in the same way as derivatives of $B$.

In the case of self-dual Yang-Mills the procedure described above is only valid at tree-level. To make it well-defined at the quantum level other fields must be added to cancel the one-loop anomaly. Fortunately this construction is generic for gauge theories, that is, it works essentially in the same way for theories with SUSY, which are anomaly free, and for non-adjoint matter by simply changing the representation of the Wilson line. The case of $\mathcal{N}=4$ super Yang-Mills is particularly nice since all the fields sit inside the same supermultiplet in twistor space, allowing the authors of \cite{Koster:2016ebi,Koster:2016loo} to give a supersymmetric generating function for the lift of all local operators in the theory.

\section{Conformal blocks from lifts of local operators}\label{blocks}

\subsection{Self-dual Yang-Mills}\label{conformalb_sdym}

With the lift of local operators to twistor space in hand, extracting its conformal blocks in the chiral algebra is quite simple. Consider again the operator $\text{Tr}\,B^2$ which was argued in \cite{Costello:2022wso} to correspond to the conformal block $\langle\text{tr}\,B^2|$ such that
\begin{multline}\label{trb2_JJJJ}
    \langle\text{tr}B^2|\tilde{J}^{\msf{a}_1}\tilde{J}^{\msf{a}_2}J^{\msf{a}_3}\cdots J^{\msf{a}_n}\rangle=\\
    \sum_{\sigma\in S_n}\,\text{tr}(T^{\sigma(a_1)}\cdots T^{\sigma(a_n)})\,\frac{z^4_{\sigma(1)\sigma(2)}}{z_{\sigma(1)\sigma(2)}z_{\sigma(2)\sigma(3)}\cdots z_{\sigma(n)\sigma(1)}}
\end{multline}
The lift to twistor space of \eqref{trb2} depends on two holomorphic Wilson lines \eqref{Wilson_line} which we expand using their definition \eqref{Wilson_line_def} to order $n$ in the field $a$. The contributions to this order are given by all the possible ways of distributing the $a$ fields between the two $g$ fields. All of these terms are given by permutations of the expression
\begin{multline}\label{B2_block}
    \int_{(\mathbb{CP})^{n}}\langle \prod_{p=1}^{n}\langle\lambda_p\d\lambda_p\rangle\frac{\langle ij\rangle^4}{\langle 12\rangle\langle 23 \rangle\langle 34 \rangle\cdots\langle n1 \rangle}\\
    \tr\,(\cdots a(\lambda_{i-1})g(\lambda_i) a(\lambda_{i+1})\cdots a(\lambda_{j-1})g(\lambda_j)a(\lambda_{j+1})\cdots)
\end{multline}
where $\la ij\ra=\epsilon^{\alpha\beta}\lambda_{i\alpha}\lambda_{j\beta}$. The field independent coefficient in the integrand is exactly the same as the correlation function for the $\langle\tr B^2|$ conformal block \eqref{trb2_JJJJ}. Furthermore, recalling that $J$ couples to the bulk field $a$ and $\tilde{J}$ couples to $g$, we can read from the fields present in \eqref{B2_block} for which chiral currents the prefactor is a correlation function of. It is clear that expanding further in powers of $a$ reproduces all the correlation functions for the conformal block $|\tr B^2\rangle$ from the lift of the local operator to twistor space.

We claim that this is valid for any local operator in self-dual Yang-Mills. Its lift to twistor space constructs explicitly all the chiral algebra correlators of its corresponding conformal block dual. The numerator is fixed by the fields representative in twistor space which can be read out of the Penrose transform while the denominator appears from the expansion of the holomorphic Wilson lines. This field independent coefficient gives the correlator for the conformal block corresponding to the local operator. The leftover fields can be used to identify which algebra elements are present in the correlator by looking at their bulk-boundary coupling on $\mathbb{CP}^1$. For example from the lift of the operator
\begin{equation}
\text{tr}\, B_{\alpha\beta}B_{\gamma\delta}B^{\alpha\beta}B^{\gamma\delta}\rightarrow \int_{(\mathbb{CP}^1)^4}\langle ik\rangle^2\langle jl\rangle^2\text{tr} \left(g_iW[\lambda_i,\lambda_j]g_jW[\lambda_j,\lambda_k]g_kW[\lambda_k,\lambda_l]g_lW[\lambda_l,\lambda_i]\right)
\end{equation}
we can read out its conformal block
\begin{multline}
\langle\tr B_{\alpha\beta}B_{\gamma\delta}B^{\alpha\beta}B^{\gamma\delta}|\cdots  \tilde{J}_i^{\msf{a}_i}\cdots\tilde{J}_j^{\msf{a}_j}\cdots\tilde{J}_k^{\msf{a}_k}\cdots\tilde{J}_l^{\msf{a}_l}\cdots\rangle=\sum_{\sigma\in S_n}\,\text{tr}(T^{\sigma(a_1)}\cdots T^{\sigma(a_n)})\,\\
\frac{\la\lambda_{\sigma(i)}\lambda_{\sigma(k)}\ra^2\la\lambda_{\sigma(j)}\lambda_{\sigma(l)}\ra^2\la\lambda_{\sigma(i)}\lambda_{\sigma(j)}\ra\la\lambda_{\sigma(j)}\lambda_{\sigma(k)}\ra\la\lambda_{\sigma(k)}\lambda_{\sigma(l)}\ra\la\lambda_{\sigma(l)}\lambda_{\sigma(i)}\ra}{\la\lambda_{\sigma(1)}\lambda_{\sigma(2)}\ra\la\lambda_{\sigma(2)}\lambda_{\sigma(3)}\ra\cdots \la\lambda_{\sigma(n)}\lambda_{\sigma(1)}\ra}
\end{multline}
where $\cdots$ contains only insertions of the chiral current $J$. 
For operators with derivatives of $B$ and $F$ the procedure is the same, the field independent coefficient of their lift to twistor space gives the 2d correlator for the currents dual to the leftover fields. In this case, derivatives of operators couple to the KK chiral currents  
\begin{equation}
\partial^m_{\mu^{\dot{0}}}\partial^n_{\mu^{\dot{1}}}g\, \tilde{J}[m,n]\; ,\;\;\partial^m_{\mu^{\dot{0}}}\partial^n_{\mu^{\dot{1}}}a\,J[m,n]
\end{equation}
where $\mu^{\dot{0}}$ and $\mu^{\dot{1}}$ are two components of $\mu^{\dot{\alpha}}$ transforming under $SL(2)$. An example containing KK currents is given by the lifted operator
\begin{equation}
\tr\,\partial_{\alpha\dot\alpha}B_{\gamma\delta}\partial^{\delta\dot\alpha}B^{\gamma}{}_{\sigma}B^{\sigma\alpha}\rightarrow \int_{(\mathbb{CP}^1)^3}\langle ij\rangle^2\langle ik\rangle\langle jk\rangle \tr\,\left(\partial_{\dot\alpha}g_iW[\lambda_i,\lambda_j]\partial^{\dot\alpha}g_jW[\lambda_j,\lambda_k]g_kW[\lambda_k,\lambda_i]\right)
\end{equation}
which corresponds to the conformal blocks
\begin{align}
&\langle \tr\,\partial_{\alpha\dot{0}}B_{\gamma\delta}\partial^{\delta\dot{0}}B^{\gamma}{}_{\sigma}B^{\sigma\alpha}|\cdots \tilde{J}^{\msf{a}_i}[1,0]\cdots\tilde{J}^{\msf{a}_j}[1,0]\cdots\tilde{J}^{\msf{a}_k}\cdots\rangle\\
&=\sum_{\sigma\in S_n}\,\text{tr}(T^{\sigma(a_1)}\cdots T^{\sigma(a_n)})\,\frac{\la\lambda_{\sigma(i)}\lambda_{\sigma(j)}\ra^3\la\lambda_{\sigma(i)}\lambda_{\sigma(k)}\ra^2\la\lambda_{\sigma(j)}\lambda_{\sigma(k)}\ra^2}{\la\lambda_{\sigma(1)}\lambda_{\sigma(2)}\ra\la\lambda_{\sigma(2)}\lambda_{\sigma(3)}\ra\cdots \la\lambda_{\sigma(n)}\lambda_{\sigma(1)}\ra}\nonumber\\
&\langle \tr\,\partial_{\alpha\dot{1}}B_{\gamma\delta}\partial^{\delta\dot{1}}B^{\gamma}{}_{\sigma}B^{\sigma\alpha}|\cdots \tilde{J}^{\msf{a_i}}[0,1]\cdots\tilde{J}^{\msf{a}_j}[0,1]\cdots\tilde{J}^{\msf{a}_k}\cdots\rangle\\
&=\sum_{\sigma\in S_n}\,\text{tr}(T^{\sigma(a_1)}\cdots T^{\sigma(a_n)})\,\frac{\la\lambda_{\sigma(i)}\lambda_{\sigma(j)}\ra^3\la\lambda_{\sigma(i)}\lambda_{\sigma(k)}\ra^2\la\lambda_{\sigma(j)}\lambda_{\sigma(k)}\ra^2}{\la\lambda_{\sigma(1)}\lambda_{\sigma(2)}\ra\la\lambda_{\sigma(2)}\lambda_{\sigma(3)}\ra\cdots \la\lambda_{\sigma(n)}\lambda_{\sigma(1)}\ra}\nonumber
\end{align}
where $\cdots$ only contains insertions of $J$. The lift of local operators gives in a sense an effective description of the 2d theory, where its degrees of freedom have been integrated out in terms of the field independent correlator.

What is behind this explicit construction is the fact that gauge theories in twistor space enjoy a larger gauge redundancy than their spacetime counterpart, allowing for gauge choices that are inaccessible in spacetime. In particular, in axial gauge $Z_*^{\bar{I}} a_{\bar{I}}=0=Z_*^{\bar{I}}g_{\bar{I}}$ the interaction vertex $g\wedge a\wedge a$ in \eqref{action_SDYM} is zero but we still expect that matrix elements of operators to be nonzero. That is, the form factor for $\tr B^2$, for example, should still give the MHV amplitude when integrated over spacetime. This is taken care of in twistor space by the lift of the local operator using the holomorphic Wilson lines which, in a sense, takes into account all the interactions in this axial gauge. Moreover, in this gauge the computation neatly separates into a purely 2d part given by the chiral algebra correlators and a 4d part which in this case gives a trivial delta function for momentum conservation.

\subsection{$\mathcal{N}=4$ super Yang-Mills}\label{N=4SYM}

There is an interesting connection of the ideas above and the twistor action for $\mathcal{N}=4$ super Yang-Mills given in \cite{Boels:2006ir}. In particular the 2d/4d correspondence in \cite{Costello:2022wso} gives a new interpretation for a peculiar non-local interaction vertex in that action. The $\mathcal{N}=4$ action in twistor space is more elegantly formulated in a supersymmetric extension of twistor space in which we introduce four fermionic coordinates $\chi^A$ with values on $\mathcal{O}(1)$. These coordinates, alongside the bosonic coordinates $Z^I$ are to be taken up to global rescalings by a non-zero complex number $r\in\mathbb{C}^*$ so that $\mathcal{Z}=(Z,\chi)\sim r(Z,\chi)$ are homogeneous coordinates on $\mathbb{CP}^{3|4}$, to obtain the supertwistor space we removed the line $\lambda=0$ from $\mathbb{CP}^{3|4}$. In this superspace the whole $\mathcal{N}=4$ supermultiplet can be represented as a single superfield 
\begin{multline}\label{n4_multiplet}
    \mathcal{A}(\mathcal{Z})=a(Z)+\psi_A(Z)\chi^A+\frac{1}{2!}\phi_{AB}(Z)\chi^A\chi^B\\
    +\frac{\varepsilon_{ABCD}}{3!}\bar\psi^A(Z)\chi^B\chi^C\chi^D+\frac{\varepsilon_{ABCD}}{4!}g(Z)\chi^A\chi^B\chi^C\chi^D
\end{multline}
of homogeneous weight zero. Fields in the $\chi^A$ expansion have different scaling weights and correspond to the different helicities of the $\mathcal{N}=4$ multiplet by the Penrose transform. The measure on twistor space can be extended to the fermionic coordinates to define the Berezinian
\begin{equation}
    D^{3|4}\mathcal{Z}=D^3Z\prod_{A=1}^4 d\chi^{A}
\end{equation}
and the Lagrangian for self-dual $\mathcal{N}=4$ sYM takes the form of a holomorphic Chern-Simons theory
\begin{equation}\label{susy_cs}
    S_{sd-sYM}=\int d^{3|4}\mathcal{Z}\wedge\left(\mA\wedge\bar\partial\mA+\frac{2}{3}\mA\wedge\mA\wedge\mA \right).
\end{equation}
Integrating out the fermions recovers the holomorphic $BF$ theory \eqref{action_SDYM} and its supersymmetric extension. Actions with less amount of supersymmetry can easily be obtained as well as actions with other kinds of matter fields, see \cite{Boels:2006ir} for details. 

To complete this action to full Yang-Mills a non-local interaction of the form
\begin{equation}\label{logdet}
    S_{int}=\int_{\mathbb{R}^{4|8}} d^{4|8}x\, \log \det(\bar\partial_{\mA}|_{L})
\end{equation}
is added, where $L$ is a linearly embedded $\mathbb{CP}^1$ in supertwistor space, $\bar\partial_{\mathcal{A}}= \bar\partial + \mathcal{A}$ is the covariant derivative for the partial connection given by the superfield $\mathcal{A}$ and the integral is over the moduli space of linearly embedded lines in supertwistor space $\mathbb{R}^{4|8}$. In practical computations this vertex is expanded  perturbatively using the Green's function of the $\bar\partial$ operator \cite{Boels:2006ir,Adamo:2013cra}
\begin{equation}\label{logdet}
    \log\det(\bar\partial_{\mathcal{A}}|_{L}) = \tr\log(\bar\partial)|_L +  \sum_{n=2}^{\infty} \frac{1}{n}\int_{L^n} \tr\left(\bar\partial^{-1}|_L\mA_1\bar\partial^{-1}|_L\mA_2\cdots\bar\partial^{-1}|_L\mA_n\right).
\end{equation}
In this form we see the connection with the lift of local operators, the holomorphic Wilson line is built out of the differential \eqref{mero_dif}, which is the inverse of the $\bar\partial$ operators\eqref{inverse}. Performing the fermionic integrals kills the first term while the second term gives the lift of several operators to twistor space. These lifted operators are built of all the possible combinations of fields with a total combined a weight of $-8$. Taking again a gauge adapted to spacetime, the only non-zero operators in \eqref{logdet} are $\tr g^2$, $\tr \phi^4$ and $\tr \phi\bar\psi\bar\psi$ which Penrose transform to $\tr B^2$, $\tr \Phi^4$ and $\tr \bar\Psi^2\Phi$ where $\Psi_{A\alpha}$ is the positive helicity gluino, $\bar\Psi^A_{\dot\alpha}$ is the negative gluino, and $\Phi_{AB}$ is the biadjoint scalar. These are the missing interaction that complete the self-dual sector to the full $\mathcal{N}=4$ Yang-Mills theory given in spacetime by the action
\begin{align}\label{SUSY_CS}
    &S_{\text{SUSY-CS}} =\nonumber \\
    &\int_{\mathbb{R}^4} d^4 x\, \tr\left(\frac{1}{2}B_{\alpha\beta}F^{\alpha\beta}+\bar{\Psi}^A{}_{\dal} D^{\alpha\dal} \Psi_{A\alpha}+ \frac{\veps^{ABCD}}{16}D^{\alpha}{}_{\dal}\Phi_{AB}D_{\alpha}{}^{\dal}\Phi_{CD}
    +\frac{\veps^{ABCD}}{2}\Phi_{AB}\Psi_C{}^{\alpha}\Psi_{D\alpha}\right) \nonumber\\
    &-\mathrm{g}\int_{\mathbb{R}^4} d^4x\, \tr\left(\frac{1}{2}B_{\alpha\beta}B^{\alpha\beta}+\frac{1}{2}\Phi_{AB}\bar\Psi^A{}_{\dal}\bar\Psi^{B\dal}+\frac{1}{16} \veps^{ACDE}\veps^{BFGH}\Phi_{AB}\Phi_{CD}\Phi_{EF}\Phi_{GH} \right)
\end{align}
However, in a generic gauge there are many other operators on the right hand side of \eqref{logdet}. 

In here lies a subtlety of the supersymmetric theory. In an axial gauge, the lift of $\tr B^2$ to twistor space gives the operator \eqref{B2_block}, whose form factor does not match the form factor for $\tr B^2$ as computed from the spacetime action. In the supersymmetric theory, axial gauge kills more than just the $g\wedge a\wedge a$ vertex, it kills all the interactions in the $\mA\wedge\mA\wedge\mA$ vertex, which are not taken into account by the lift using the purely bosonic Wilson line. On the other hand the operators contained in the vertex \eqref{logdet} are enough to reconstruct the complete form factor for $\tr B^2$. This suggests that local operators have to be lifted in a supersymmetric form to twistor space. This issue was first pointed out in \cite{Koster:2016ebi,Koster:2016loo} where a prescription for the lift for all local operators in $\mathcal{N}=4$ sYM to twistor space was given using a supersymmetric prescription. This might seem at first to leave us with a conundrum, in the supersymmetric theory taking the naive lift doesn't give the correct operator. On the other hand, using the supersymmetric lift gives too many operators, producing a plethora of conformal blocks according to our prescription. We take the viewpoint that this is not an issue, the naive lift produces one conformal block per operator irrespective of the amount of supersymmetry. The fact that the naive lifted operator is in a sense incomplete in the supersymmetric theory is not an issue if we are only interested in the dual conformal block, which is still uniquely defined. The arguments in \cite{Costello:2022wso} don't require an actual operator in twistor space to be constructed for the 2d/4d duality to hold. It is nevertheless interesting to consider what the natural interpretation of the supersymmetric lift of local operators is in terms of a possible supersymmetric 2d/4d correspondence and how this fits into the topological string construction of the self-dual action on twistor space of \cite{Witten:2003nn}. We hope to address this in upcoming work.

With this caveat in mind, the prescription in \ref{conformalb_sdym} is still valid for all the local operators in $\mathcal{N}=4$ sYM.  We could also lift the local operators contained in the log det interaction vertex using the bosonic Wilson lines. In \eqref{SUSY_CS}, other than the $\tr B^2$ operator we have already seen in the self-dual case, we could also have $\tr\Phi^4$ and $\tr\Psi^2\Phi$, they both reside on the same line on twistor space. For example, $\tr\Phi^4(x)$ can be lifted to twistor space in the following way
\begin{multline}
    \tr \Phi^4(x) \rightarrow  \int_{(\mathbb{CP}^1)^4}\,\la\lambda_i\d\lambda_i\ra\la\lambda_j\d\lambda_j\ra\la\lambda_k\d\lambda_k\ra\la\lambda_l\d\lambda_l\ra\,\veps^{ACDE}\veps^{BFGH} \tr\left(\phi_{AB}(x\lambda_i,\lambda_i)W[\lambda_i,\lambda_j] \right.\\
   \left. \phi_{CD}(x\lambda_j,\lambda_j)W[\lambda_j,\lambda_k] \phi_{EF}(x\lambda_k,\lambda_k)W[\lambda_k,\lambda_l]
    \phi_{GH}(x\lambda_l,\lambda_l) W[\lambda_l,\lambda_i]\right)
\end{multline}
It is straightforward to expand the Wilson lines explicitly into integrals over copies of $\mathbb{CP}^1$s as in \eqref{B2_block} and read out the conformal block in this case
\begin{multline}\label{D4_Block}
    \la\text{tr} \Phi^4(x) \vert \veps_{ACDE}\veps_{BFGH} J_{a}^{\msf{a}_1}(z_1)\cdots J_{\phi}^{\msf{a}_i,AB}(z_i)\cdots J_{\phi}^{\msf{a}_j,CD}(z_j)\cdots J_{\phi}^{\msf{a}_k,EF}(z_k)\cdots
    J_{\phi}^{\msf{a}_l,GH}(z_l)\cdots J_{a}^{\msf{a}_n}(z_n)\ra  \\
    =\sum_{\sigma\in S_n}\,\text{tr}(T^{\sigma(a_1)}\cdots T^{\sigma(a_n)})\,\frac{\la\lambda_{\sigma(i)}\lambda_{\sigma(j)}\ra\la\lambda_{\sigma(j)}\lambda_{\sigma(k)}\ra\la\lambda_{\sigma(k)}\lambda_{\sigma(l)}\ra\la\lambda_{\sigma(l)}\lambda_{\sigma(i)}\ra}{\la\lambda_{\sigma(1)}\lambda_{\sigma(2)}\ra\la\lambda_{\sigma(2)}\lambda_{\sigma(3)}\ra\cdots \la\lambda_{\sigma(n)}\lambda_{\sigma(1)}\ra}
\end{multline}
where we have used $J_{a}^{\msf{a}}$ to label the dual chiral current that couples to $a$ on $\mathbb{CP}^1$ and $J_{\phi}^{\msf{a},AB}$ to label the dual current for $\phi_{AB}$. Just as before, the dots in between different $J_{\phi}^{\msf{a},AB}$ currents are insertions of $J_{a}^{\msf{a}}$ currents. For example, permutations of terms such as $\la ij\ra^2\la kl\ra^2$, $\la ij\ra\la jk\ra\la kl\ra\la li\ra$ will appear in the sum. This appearance of multiple terms in the conformal block is not surprising, we expect the same result to be produced using correlators of the chiral currents written in \eqref{D4_Block}, reducing the correlator to known two point correlators using OPEs between chiral currents could be achieved in different ways, we shall see this in \ref{current_sym}.

We could also look at the bosonic lift of $\tr \bar\Psi^2\Phi(x)$ to twistor space
\begin{multline}
    \tr \bar\Psi^2\Phi(x) \rightarrow  \int_{(\mathbb{CP}^1)^3}\,\la\lambda_i\d\lambda_i\ra\la\lambda_j\d\lambda_j\ra\la\lambda_k\d\lambda_k\ra\la jk\ra\, \tr\left(\phi_{AB}(x\lambda_i,\lambda_i)W[\lambda_i,\lambda_j] \right.\\
   \left. \bar\psi^{A}(x\lambda_j,\lambda_j)W[\lambda_j,\lambda_k] \bar\psi^B(x\lambda_k,\lambda_k)W[\lambda_k,\lambda_i]\right)
\end{multline}
Expanding the Wilson lines should give us the following integrals over $n$ copies of $\mathbb{CP}^1$
\begin{multline}
    \int_{(\mathbb{CP})^{n}}\langle \prod_{p=1}^{n}\langle\lambda_p\d\lambda_p\rangle\frac{\la jk\ra^2\la ij\ra\la ki\ra}{\langle 12\rangle\langle 23 \rangle\langle 34 \rangle\cdots\langle n1 \rangle}\,\la ij\ra\la jk\ra\la ki\ra\\
    \tr\,(\cdots a(\lambda_{i-1})\phi_{AB}(\lambda_i) a(\lambda_{i+1})\cdots a(\lambda_{j-1})\bar\psi^A(\lambda_j)a(\lambda_{j+1})\cdots a(\lambda_{k-1})\bar\psi^B(\lambda_k)a(\lambda_{k+1})\cdots)
\end{multline}
hence we write the conformal block corresponding to the local operator $\tr \bar\Psi^2\Phi(x)$ as 
\begin{multline}\label{C^2D_Block}
    \la\tr \bar\Psi^2\Phi(x) \vert  J_{a}^{\msf{a}_1}(z_1)\cdots J_{\phi}^{\msf{a}_i,AB}(z_i)\cdots J_{\bar\psi,A}^{\msf{a}_j}(z_j)\cdots J_{\bar\psi,B}^{\msf{a}_k}(z_k)\cdots
   J_{a}^{\msf{a}_n}(z_n)\ra \\
    = \sum_{\sigma\in S_n}\,\text{tr}(T^{\sigma(a_1)}\cdots T^{\sigma(a_n)})\,   \frac{\la\lambda_{\sigma(i)}\lambda_{\sigma(j)}\ra^2\la\lambda_{\sigma(j)}\lambda_{\sigma(k)}\ra^3\la\lambda_{\sigma(k)}\lambda_{\sigma(i)}\ra^2}{\la\lambda_{\sigma(1)}\lambda_{\sigma(2)}\ra\la\lambda_{\sigma(2)}\lambda_{\sigma(3)}\ra\cdots \la\lambda_{\sigma(n)}\lambda_{\sigma(1)}\ra}
\end{multline}
where we have used $J_{\bar\psi,A}^{\msf{a}}$ to denote the chiral current that couples to the gluino $\bar\psi^A$. So far we have seen examples of conformal blocks of local operators inside the log det vertex \eqref{logdet}, one could easily apply the bosonic lifting procedure to other local operators in the theory, for example the operator $\tr\,\Phi^2$ lifts to 
\begin{multline}
    \tr \,\Phi^2(x) \rightarrow  \int_{(\mathbb{CP}^1)^2}\,\la\lambda_i\d\lambda_i\ra\la\lambda_j\d\lambda_j\ra\,\veps^{ABCD} \tr\left(\phi_{AB}(x\lambda_i,\lambda_i)W[\lambda_i,\lambda_j] \right.\\
   \left. \phi_{CD}(x\lambda_j,\lambda_j)W[\lambda_j,\lambda_i]\right)
\end{multline}
and from the expansion of the Wilson lines we find the conformal block
\begin{multline}
    \la\tr\, \Phi^2 \vert \veps_{ABCD} J_{a}^{\msf{a}_1}(z_1)\cdots J_{\phi}^{\msf{a}_i,AB}(z_i)\cdots J_{\phi}^{\msf{a}_j,CD}(z_j)\cdots J_{a}^{\msf{a}_n}(z_n)\ra \\
    = \sum_{\sigma\in S_n}\,\text{tr}(T^{\sigma(a_1)}\cdots T^{\sigma(a_n)})\,   \frac{\la\lambda_{\sigma(i)}\lambda_{\sigma(j)}\ra^2}{\la\lambda_{\sigma(1)}\lambda_{\sigma(2)}\ra\la\lambda_{\sigma(2)}\lambda_{\sigma(3)}\ra\cdots \la\lambda_{\sigma(n)}\lambda_{\sigma(1)}\ra}
\end{multline}
Similarly, the local operator $\tr\,(\partial_{\alpha\dal} \Phi\partial^{\alpha\dal}\Phi)$ lifts to
\begin{multline}
    \tr\,(\partial \Phi)^2(x) \rightarrow  \int_{(\mathbb{CP}^1)^2}\,\la\lambda_i\d\lambda_i\ra\la\lambda_j\d\lambda_j\ra\la ij\ra\,\veps^{ABCD} \tr\left(\partial_{\mu^{\dal}}\phi_{AB}(x\lambda_i,\lambda_i)W[\lambda_i,\lambda_j] \right.\\
   \left. \partial_{\mu_{\dal}}\phi_{CD}(x\lambda_j,\lambda_j)W[\lambda_j,\lambda_i]\right)
\end{multline}
from which we obtain the conformal blocks
\begin{align}
    \la\tr\, &\partial_{\alpha\dot{0}} \Phi\partial^{\alpha\dot{0}} \Phi \vert \veps_{ABCD} J_{a}^{\msf{a}_1}(z_1)\cdots J_{\phi}^{\msf{a}_i,AB}[1,0](z_i)\cdots J_{\phi}^{\msf{a}_j,CD}[1,0](z_j)\cdots J_{a}^{\msf{a}_n}(z_n)\ra \nonumber
    \\& = \sum_{\sigma\in S_n}\,\text{tr}(T^{\sigma(a_1)}\cdots T^{\sigma(a_n)})\,   \frac{\la\lambda_{\sigma(i)}\lambda_{\sigma(j)}\ra^3}{\la\lambda_{\sigma(1)}\lambda_{\sigma(2)}\ra\la\lambda_{\sigma(2)}\lambda_{\sigma(3)}\ra\cdots \la\lambda_{\sigma(n)}\lambda_{\sigma(1)}\ra} 
    \\\la\tr\, &\partial_{\alpha\dot{1}} \Phi\partial^{\alpha\dot{1}} \Phi \vert \veps_{ABCD} J_{a}^{\msf{a}_1}(z_1)\cdots J_{\phi}^{\msf{a}_i,AB}[0,1](z_i)\cdots J_{\phi}^{\msf{a}_j,CD}[0,1](z_j)\cdots J_{a}^{\msf{a}_n}(z_n)\ra \nonumber
    \\&= \sum_{\sigma\in S_n}\,\text{tr}(T^{\sigma(a_1)}\cdots T^{\sigma(a_n)})\,   \frac{\la\lambda_{\sigma(i)}\lambda_{\sigma(j)}\ra^3}{\la\lambda_{\sigma(1)}\lambda_{\sigma(2)}\ra\la\lambda_{\sigma(2)}\lambda_{\sigma(3)}\ra\cdots \la\lambda_{\sigma(n)}\lambda_{\sigma(1)}\ra} 
\end{align}
with $J_{\phi}^{\msf{a}_i,AB}[m,n]$ chiral current dual to the KK modes of the biadjoint scalar.

\section{Representation of chiral algebras}\label{rep_chiral_algebra}
\subsection{Self-dual Yang-Mills}
We have seen that the procedure of lifting local operators from spacetime to twistor space automatically construct the correlation functions for the conformal block associated to the local operator by the 2d/4d correspondence. In this way of proceeding, the degrees of freedom on the lines have already been integrated out, leaving only the explicit separation between the 2d part of the computation and the 4d part. In this section we will see that it is straightforward to give explicit realizations for some of these degrees of freedom in terms of free chiral fields in a 2d CFT. We don't claim that these are the actual microscopic degrees of freedom, but they are enough to reproduce the computation of the conformal blocks as correlation functions of a 2d CFT and may be useful in exploring generalizations. The guiding principle will be once again gauge invariance but in a fashion similar to \cite{Costello:2022wso}.

The chiral algebra lives on a line in twistor space and should have a gauge invariant coupling to the bulk fields. 
The coupling of the positive helicity field $a$ to the chiral current $J^{\msf{a}}$ takes the form 
\begin{equation}
\int_{\mathbb{CP}^1} \tr (J\, a(z))
\end{equation}
with $a(z)$ the pullback to $\mathbb{CP}^1$ of the partial connection. This coupling only makes sense if $J^{\msf{a}}$ is a section of the canonical bundle $K$ over $\mathbb{CP}^1$, so that it can be integrated over the sphere. We must be able to exponentiate this coupling and add it to an action describing the interaction between bulk and boundary fields. Requiring gauge invariance of this exponentiation enforces the OPE \eqref{ca_sdYM}, as well as the vanishing of its level. To represent this in a 2d CFT we take $J^\msf{a}$  to be a Kac-Moody current $j^\msf{a}$ and restrict ourselves to the single trace sector. The coupling to the negative helicity field $g$ is similar
\begin{equation}
\int_{\mathbb{CP}^1}\tr(\tilde{J}g(z))
\end{equation}
with $\tilde{J}^{\msf{a}}$ also a section of the canonical bundle on $\mathbb{CP}^1$, but the weight of $g$ has to be taken into account. To do this we write $\tilde{J}^{\msf{a}}=\mathcal{O}\,j^{\msf{a}}$ as a product of the same Kac-Moody current $j^{\msf{a}}$ as before and another operator $\mathcal{O}$. This operator has to cancel the weight of $g$ which is a section of $\mathcal{O}(-4)$ restricted to $\mathbb{CP}^1$. This bundle is equivalent to $K^2$\footnote{Sections of $\mathcal{O}(-2)$ have the same transition functions as sections of $K$ over $\mathbb{CP}^1$, thus they are identified as line bundles.} so the operator $\mathcal{O}$ should take values in the square of the holomorphic tangent bundle $T^2$. Gauge invariance of this interaction also requires that $\tilde{J}\tilde{J}\sim0$, so the $\mathcal{O}(z)\mathcal{O}(w)$ OPE should start at most at order $(z-w)$. These constraints can be solved using a fermionic beta-gamma system with a fermion $\eta$ with values in $T^{3/2}$ such that $\tilde{J}^{\msf{a}}=\eta\partial\eta \,j^{\msf{a}}$. 

With these definitions of $J^{\msf{a}}$ and $\tilde{J}^{\msf{a}}$ it is simple to check that the OPEs \eqref{ca_sdYM} and \eqref{ca_+-} are reproduced. Moreover, there is a selection rule for the correlation functions of these operators due to the 4 fermionic zero modes of $\eta$ on the sphere. This implies that the only non-vanishing correlation functions are those with exactly two insertions of $\tilde{J}^{\msf{a}}$ since each absorbs two zero modes. A quick computation yields
\begin{multline}
\langle\tilde{J}_1^{\msf{a}_1}(z_1)\tilde{J}_2^{\msf{a}_2}(z_2) J_3^{\msf{a}_3}(z_3) \cdots J_n^{\msf{a}_n}(z_n) \rangle=\tr\left(T^{\msf{a}_1}\cdots T^{\msf{a}_n}\right) \left\langle\prod_{i=1}^n j_n\right\rangle\left\langle (\eta\partial\eta)_1(\eta\partial\eta)_2\right\rangle\\=\tr\left(T^{\msf{a}_1}\cdots T^{\msf{a}_n}\right) \frac{\langle 12\rangle^4}{\langle 12\rangle\langle 23\rangle\cdots\langle n1\rangle}  +\dots
\end{multline}
where we only wrote explicitly one color ordering and kept only the single trace contributions. The denominator comes from saturating the zero modes of $\eta$ which are the global holomorphic sections of $T^{3/2}$. We can parametrize the zero-modes by $\eta_{\text{zm}}(z)=\eta_0 +z\eta_1+z^2\eta_2+z^3\eta_3$, then it is easy to performs the finite fermionic integral over $\eta_i$'s producing the numerator.

It is interesting that the simplest way to couple the bulk and boundary theories satisfying gauge invariance naturally produces the conformal block for the operator $\tr B^2$. For higher polynomials $\eta$ can be twisted\footnote{To preserve invariance under rescaling, the restriction of $\mathcal{O}(-4)$ must also be twisted by powers of $K$.} by of $T^{n}$ providing more zero modes that are absorbed by more insertions of $\tilde{J}$. The correlation functions then correspond to all possible ways of contracting the indices of $B$ to construct higher degree polynomials out of them. 



\subsection{$\mathcal{N}=4$ sYM}\label{current_sym}

We can proceed in a similar fashion for the $\mathcal{N}=4$ supermultiplet. Each bulk field in~\eqref{n4_multiplet} couples to a chiral field living on the line defect
\begin{equation}
\begin{array}{ccccc}
    \int \tr(J_{a} \,a),\;  & \int \tr(J_{\psi}^{A} \,\psi_A),\; & \int \tr (J_{\Phi}^{AB} \,\Phi_{AB}),\; & \int \tr(J_{\bar\psi,D}\,\bar\psi^D),\; & \int \tr( J_{g}\,g)
    \end{array}
\end{equation}
Enforcing gauge invariance in the same manner as before results in same OPEs as~\eqref{ca_sdYM} and \eqref{ca_+-} for $J_a$ and $J_g$ as well as requiring that all the other currents transform in the adjoint and have regular OPEs among themselves. This is easily achieved by once again taking the single trace sector of a Kac-Moody current $j$ to represent $J_a$ and and taking all other currents to be given by some operator $\mathcal{O}^R$ carrying R-symmetry indices multiplied by $j$. These operators must also cancel the weights of various powers of $\mathcal{O}(-1)$ restricted to the line defect. This is taken care of by introducing four fermionic $\beta-\gamma$ systems with each fermionic field $\eta^A$ taking values in $T^{1/2}$. The representation for the remaining fields in the chiral algebra is
\begin{equation}
J_{\psi}^{\msf{a},A}=\eta^A j^{\msf{a}},\;J_\phi^{\msf{a},AB} =\eta^A\eta^B j^\msf{a},\; J_{\bar\psi,D}^{\msf{a}}=\veps_{ABCD}\eta^A\eta^B\eta^C j^{\msf{a}},\;J^{\msf{a}}_g=\epsilon_{ABCD}\eta^A\eta^B\eta^C\eta^D j^\msf{a}.
\end{equation}
Each field $\eta$ has two zero modes given by global holomorphic sections of $T^{1/2}$. This requires that non zero correlators have two insertions of each field. One way to saturate these zero modes is with two insertions of $\tilde{J}_g$ reproducing the results of the previous section for the the conformal block of $\tr B^2$. Using the other fields in the supermultiplet there are many other ways of satisfying the zero mode constraint. Not surprisingly, the non-zero correlation functions reproduce the conformal blocks of the operators that appear after performing the fermionic integrations in the log det interaction vertex \eqref{logdet}. For example the correlators
\begin{align}
    &\left\langle J_{a}^{\msf{a}_1}(z_1)\cdots J_{\phi}^{\msf{a}_i,AB}(z_i)\cdots J_{\bar\psi,A}^{\msf{a}_j}(z_j)\cdots J_{\bar\psi,B}^{\msf{a}_k}(z_k)\cdots
    J_{a}^{\msf{a}_n}(z_n) \right\rangle\nonumber \\ 
    =& \left\langle\prod_{\delta=1}^n j^{\msf{a}_{\delta}}(z_{\delta})\right\rangle\left\langle \eta^A(z_i)\eta^B(z_i)\veps_{LMHA}\eta^{L}(z_j)\eta^{M}(z_j)\eta^H(z_j)\veps_{EFGB}\eta^{E}(z_k)\eta^{F}(z_k)\eta^G(z_k)\right\rangle\nonumber\\ 
    =&\tr\left(T^{\msf{a}_1}\cdots T^{\msf{a}_n}\right) \frac{\la jk\ra^2\la ij\ra\la ki\ra}{\langle 12\rangle\langle 23\rangle\cdots\langle n1\rangle}  +\dots
\end{align}
are given by the conformal block of $\tr \bar\Psi^2\Phi$, while the conformal block for $\tr\, \Phi^4$ is given by the correlators 
\begin{align}
    &\left\langle \,\veps_{ACDE}\veps_{BFGH} J_{a}^{\msf{a}_1}(z_1)\cdots J_{\phi}^{\msf{a}_i,AB}(z_i)\cdots J_{\phi}^{\msf{a}_j,CD}(z_j)\cdots J_{\phi}^{\msf{a}_k,EF}(z_k)\cdots
    J_{\phi}^{\msf{a}_l,GH}(z_l)\cdots J_{a}^{\msf{a}_n}(z_n) \right\rangle \nonumber\\ 
    =& \left\langle\prod_{\delta=1}^n j^{\msf{a}_{\delta}}(z_{\delta})\right\rangle\left\langle\veps_{ACDE}\veps_{BFGH} \eta^A(z_i)\eta^B(z_i)\eta^C(z_j)\eta^D(z_j)\eta^{E}(z_k)\eta^{F}(z_k)\eta^G(z_l)\eta^H(z_l)\right\rangle\nonumber\\ 
    =&\tr\left(T^{\msf{a}_1}\cdots T^{\msf{a}_n}\right) \frac{\la ij\ra\la jk\ra\la kl\ra\la li\ra}{\langle 12\rangle\langle 23\rangle\cdots\langle n1\rangle}  +\dots 
\end{align}
Hence we have given an explicit realisation of the 2d chiral algebra with physical constraints. The readers familiar with the twistor string~\cite{Witten:2003nn,Berkovits:2004jj} would have noticed that this CFT is identical to a subsector of the twistor string worldsheet CFT. This identification is only valid in the MHV sector where the string worldsheet can be mapped to the line defect. At higher MHV degrees, that is higher polynomials in $B$, the worldsheet is mapped to higher degree curves while the fields of the chiral algebras still live on the linearly embedded defect line. Interestingly, the conformal blocks for $\tr B^3$ can be written in terms of squares of conformal blocks for $\tr B^2$ which is reminiscent of the CSW rules for computing amplitudes \cite{Cachazo_2004}.

Comparing to the chiral algebras of the self-dual case, we see again that the natural construction in a supersymmetric theory gives more than just the conformal block for one operator. While the CFT for the pure gluon sector in the previous section gave a realization of the chiral algebra related to $\tr B^2$, the CFT for $\mathcal{N}=4$ represents the chiral algebra for several operators related by scaling weight. Twisting the fields $\eta$ by $T^{n/2}$ give the correlation functions of other combinations of operators that can soak up all the zero modes.

\section{Discussion}\label{discussion}

We have shown how the lift of local operators to twistor space using the holomorphic Wilson line realizes explicitly the connection between 4d local operators and conformal blocks of a 2d chiral algebra. The lift to twistor space explicitly constructs all the correlation functions for the associated  conformal blocks. Although this construction applies to any gauge theory in twistor space there can be operators that fall outside its purview. For example \cite{Costello:2022wso} adds a twistor field $\eta\in\Omega^{2,1}(\mathbb{PT})$ coupling to the partial connection as
\begin{equation}
\int\eta\wedge a\wedge\partial a
\end{equation}
which is used to cancel the one-loop anomaly. To understand its degrees of freedom on spacetime requires the use of local twistors and falls outside the usual Penrose transform we employed \cite{Penrose:1986ca}. One of its degrees of freedom in spacetime is a scalar that cancels the anomaly by a Green-Schwarz mechanism, whose conformal blocks are related to the one-loop all-plus amplitude. The scalar itself is not a well-defined operator in the 4D theory due to its non-standard origin in twistor space, only its derivatives are well-defined operators. It would be interesting to develop the technology to handle operators like $\eta$ and other fields that can appear in the twisted string constructions that inspired this coupling \cite{Costello:2018zrm,Costello:2021kiv}. Although we haven't yet been able to compute the conformal block for this operator, the one-loop all-plus amplitude \cite{Bern:1993qk,Mahlon:1993si} is also supported on a line in twistor space \cite{Cachazo:2004zb}, and can be written in an interesting way using the holomorphic Wilson line
\begin{multline}
  A_\text{one-loop}(++\cdots +)=\sum_{i<j<k<l} \int_{(\mathbb{CP}^1)^4} \frac{\la\lambda_i\d\lambda_i\ra\la\lambda_j\d\lambda_j\ra\la\lambda_k\d\lambda_k\ra\la\lambda_l\d\lambda_l\ra}{\la ij\ra\la jk\ra\la kl\ra\la li\ra}\\   \lambda_{i\alpha} 
    \frac{\partial a_i}{\partial\mu_i^{\dal}} W[\lambda_i,\lambda_j] \lambda_{j\beta} \frac{\partial a_j}{\partial\mu_{j,\dal}} W[\lambda_j,\lambda_k] \lambda_{k\beta}
    \frac{\partial a_k}{\partial\mu_k^{\Dot{\beta}}} W[\lambda_k,\lambda_l] \lambda_{l\alpha}  \frac{\partial a_l}{\partial\mu_{l,\Dot{\beta}}} W[\lambda_l,\lambda_i].
\end{multline}
It resembles the generating functional in \cite{Boels:2007gv}, but neither are the lift of local operators from spacetime. It would be interesting to understand the origin of these expressions and their connection with the field $\eta$. 

Another interesting direction is to find the generalization to gravitational theories. Self-dual gravity can also be described in twistor space by a holomorphic $BF$ type action in which the Poisson structure plays a key role. The 4d/2d correspondence should also hold for self-dual gravity, and we expect that a construction similar to the gauge theory one here to be true. But instead of the usual gauge group the Wilson line should generate the parallel transport for the Poisson structure along the sphere, linking operators at different points. Moreover it might explain the appearance of the field $\eta$ and how it couples the gravitational and gauge sectors.

Lastly, we draw attention to the fact that all the statements about the twistor actions and their relation to spacetime theories should also hold true when perturbatively expanded around self-dual backgrounds. Heuristically, in these backgrounds only the dotted spinors are curved, while the undotted ones remain the same. Amplitudes in these backgrounds still have MHV prefactors \cite{Adamo:2020yzi,Adamo:2022mev} which foreshadows the existence of a 4d/2d correspondence continues to hold in certain curved backgrounds.

\section*{Acknowledgements}

We are grateful to Tim Adamo, Simon Heuveline, David Skinner and Atul Sharma for insightful conversations. We also thank Tim Adamo for commenting on the script. WB is supported by Royal Society Studentship, EC is supported by the Frankel-Goldfield-Valani Research Fund.


\bibliographystyle{JHEP}
\bibliography{cope}

\providecommand{\href}[2]{#2}\begingroup\raggedright\begin{thebibliography}{10}

\bibitem{He:2014laa}
T.~He, V.~Lysov, P.~Mitra, and A.~Strominger, {\it {BMS supertranslations and
  Weinberg\textquoteright{}s soft graviton theorem}},  {\em JHEP} {\bf 05}
  (2015) 151, [\href{http://arxiv.org/abs/1401.7026}{{\tt arXiv:1401.7026}}].

\bibitem{Strominger:2013jfa}
A.~Strominger, {\it {On BMS Invariance of Gravitational Scattering}},  {\em
  JHEP} {\bf 07} (2014) 152, [\href{http://arxiv.org/abs/1312.2229}{{\tt
  arXiv:1312.2229}}].

\bibitem{He:2014cra}
T.~He, P.~Mitra, A.~P. Porfyriadis, and A.~Strominger, {\it {New Symmetries of
  Massless QED}},  {\em JHEP} {\bf 10} (2014) 112,
  [\href{http://arxiv.org/abs/1407.3789}{{\tt arXiv:1407.3789}}].

\bibitem{He:2015zea}
T.~He, P.~Mitra, and A.~Strominger, {\it {2D Kac-Moody Symmetry of 4D
  Yang-Mills Theory}},  {\em JHEP} {\bf 10} (2016) 137,
  [\href{http://arxiv.org/abs/1503.02663}{{\tt arXiv:1503.02663}}].

\bibitem{Kapec:2014opa}
D.~Kapec, V.~Lysov, S.~Pasterski, and A.~Strominger, {\it {Semiclassical
  Virasoro symmetry of the quantum gravity $ \mathcal{S}$-matrix}},  {\em JHEP}
  {\bf 08} (2014) 058, [\href{http://arxiv.org/abs/1406.3312}{{\tt
  arXiv:1406.3312}}].

\bibitem{Kapec:2016jld}
D.~Kapec, P.~Mitra, A.-M. Raclariu, and A.~Strominger, {\it {2D Stress Tensor
  for 4D Gravity}},  {\em Phys. Rev. Lett.} {\bf 119} (2017), no.~12 121601,
  [\href{http://arxiv.org/abs/1609.00282}{{\tt arXiv:1609.00282}}].

\bibitem{Pasterski:2016qvg}
S.~Pasterski, S.-H. Shao, and A.~Strominger, {\it {Flat Space Amplitudes and
  Conformal Symmetry of the Celestial Sphere}},  {\em Phys. Rev. D} {\bf 96}
  (2017), no.~6 065026, [\href{http://arxiv.org/abs/1701.00049}{{\tt
  arXiv:1701.00049}}].

\bibitem{Campiglia:2015kxa}
M.~Campiglia and A.~Laddha, {\it {Asymptotic symmetries of gravity and soft
  theorems for massive particles}},  {\em JHEP} {\bf 12} (2015) 094,
  [\href{http://arxiv.org/abs/1509.01406}{{\tt arXiv:1509.01406}}].

\bibitem{Strominger:2013lka}
A.~Strominger, {\it {Asymptotic Symmetries of Yang-Mills Theory}},  {\em JHEP}
  {\bf 07} (2014) 151, [\href{http://arxiv.org/abs/1308.0589}{{\tt
  arXiv:1308.0589}}].

\bibitem{Campiglia:2015qka}
M.~Campiglia and A.~Laddha, {\it {Asymptotic symmetries of QED and
  Weinberg\textquoteright{}s soft photon theorem}},  {\em JHEP} {\bf 07} (2015)
  115, [\href{http://arxiv.org/abs/1505.05346}{{\tt arXiv:1505.05346}}].

\bibitem{Pasterski:2017kqt}
S.~Pasterski and S.-H. Shao, {\it {Conformal basis for flat space amplitudes}},
   {\em Phys. Rev. D} {\bf 96} (2017), no.~6 065022,
  [\href{http://arxiv.org/abs/1705.01027}{{\tt arXiv:1705.01027}}].

\bibitem{Strominger:2014pwa}
A.~Strominger and A.~Zhiboedov, {\it {Gravitational Memory, BMS
  Supertranslations and Soft Theorems}},  {\em JHEP} {\bf 01} (2016) 086,
  [\href{http://arxiv.org/abs/1411.5745}{{\tt arXiv:1411.5745}}].

\bibitem{Ball:2019atb}
A.~Ball, E.~Himwich, S.~A. Narayanan, S.~Pasterski, and A.~Strominger, {\it
  {Uplifting AdS$_{3}$/CFT$_{2}$ to flat space holography}},  {\em JHEP} {\bf
  08} (2019) 168, [\href{http://arxiv.org/abs/1905.09809}{{\tt
  arXiv:1905.09809}}].

\bibitem{He:2019jjk}
T.~He and P.~Mitra, {\it {Asymptotic symmetries and Weinberg\textquoteright{}s
  soft photon theorem in Mink$_{d+2}$}},  {\em JHEP} {\bf 10} (2019) 213,
  [\href{http://arxiv.org/abs/1903.02608}{{\tt arXiv:1903.02608}}].

\bibitem{Henneaux:2019yqq}
M.~Henneaux and C.~Troessaert, {\it {Asymptotic structure of electromagnetism
  in higher spacetime dimensions}},  {\em Phys. Rev. D} {\bf 99} (2019), no.~12
  125006, [\href{http://arxiv.org/abs/1903.04437}{{\tt arXiv:1903.04437}}].

\bibitem{Guevara:2021abz}
A.~Guevara, E.~Himwich, M.~Pate, and A.~Strominger, {\it {Holographic symmetry
  algebras for gauge theory and gravity}},  {\em JHEP} {\bf 11} (2021) 152,
  [\href{http://arxiv.org/abs/2103.03961}{{\tt arXiv:2103.03961}}].

\bibitem{Strominger:2021mtt}
A.~Strominger, {\it {$w_{1+\infty}$ Algebra and the Celestial Sphere: Infinite
  Towers of Soft Graviton, Photon, and Gluon Symmetries}},  {\em Phys. Rev.
  Lett.} {\bf 127} (2021), no.~22 221601.

\bibitem{Penrose:1967wn}
R.~Penrose, {\it {Twistor algebra}},  {\em J. Math. Phys.} {\bf 8} (1967) 345.

\bibitem{Adamo:2021lrv}
T.~Adamo, L.~Mason, and A.~Sharma, {\it {Celestial $w_{1+\infty}$ Symmetries
  from Twistor Space}},  {\em SIGMA} {\bf 18} (2022) 016,
  [\href{http://arxiv.org/abs/2110.06066}{{\tt arXiv:2110.06066}}].

\bibitem{Adamo_2022}
T.~Adamo, W.~Bu, E.~Casali, and A.~Sharma, {\it {Celestial operator products
  from the worldsheet}},  {\em JHEP} {\bf 06} (2022) 052,
  [\href{http://arxiv.org/abs/2111.02279}{{\tt arXiv:2111.02279}}].

\bibitem{Bu:2021avc}
W.~Bu, {\it {Supersymmetric celestial OPEs and soft algebras from the
  ambitwistor string worldsheet}},  {\em Phys. Rev. D} {\bf 105} (2022), no.~12
  126029, [\href{http://arxiv.org/abs/2111.15584}{{\tt arXiv:2111.15584}}].

\bibitem{Ward:1977ta}
R.~S. Ward, {\it {On Selfdual gauge fields}},  {\em Phys. Lett. A} {\bf 61}
  (1977) 81--82.

\bibitem{Mason:2005zm}
L.~J. Mason, {\it {Twistor actions for non-self-dual fields: A Derivation of
  twistor-string theory}},  {\em JHEP} {\bf 10} (2005) 009,
  [\href{http://arxiv.org/abs/hep-th/0507269}{{\tt hep-th/0507269}}].

\bibitem{Boels:2006ir}
R.~Boels, L.~J. Mason, and D.~Skinner, {\it {Supersymmetric Gauge Theories in
  Twistor Space}},  {\em JHEP} {\bf 02} (2007) 014,
  [\href{http://arxiv.org/abs/hep-th/0604040}{{\tt hep-th/0604040}}].

\bibitem{Mason:2007ct}
L.~J. Mason and M.~Wolf, {\it {Twistor Actions for Self-Dual Supergravities}},
  {\em Commun. Math. Phys.} {\bf 288} (2009) 97--123,
  [\href{http://arxiv.org/abs/0706.1941}{{\tt arXiv:0706.1941}}].

\bibitem{Witten:2003nn}
E.~Witten, {\it {Perturbative gauge theory as a string theory in twistor
  space}},  {\em Commun. Math. Phys.} {\bf 252} (2004) 189--258,
  [\href{http://arxiv.org/abs/hep-th/0312171}{{\tt hep-th/0312171}}].

\bibitem{Berkovits:2004jj}
N.~Berkovits and E.~Witten, {\it {Conformal supergravity in twistor-string
  theory}},  {\em JHEP} {\bf 08} (2004) 009,
  [\href{http://arxiv.org/abs/hep-th/0406051}{{\tt hep-th/0406051}}].

\bibitem{Skinner:2013xp}
D.~Skinner, {\it {Twistor strings for $ \mathcal{N} $ = 8 supergravity}},  {\em
  JHEP} {\bf 04} (2020) 047, [\href{http://arxiv.org/abs/1301.0868}{{\tt
  arXiv:1301.0868}}].

\bibitem{Costello:2022wso}
K.~Costello and N.~M. Paquette, {\it {Celestial holography meets twisted
  holography: 4d amplitudes from chiral correlators}},
  \href{http://arxiv.org/abs/2201.02595}{{\tt arXiv:2201.02595}}.

\bibitem{Costello:2018zrm}
K.~Costello and D.~Gaiotto, {\it {Twisted Holography}},
  \href{http://arxiv.org/abs/1812.09257}{{\tt arXiv:1812.09257}}.

\bibitem{Costello:2020jbh}
K.~Costello and N.~M. Paquette, {\it {Twisted Supergravity and Koszul Duality:
  A case study in AdS$_3$}},  {\em Commun. Math. Phys.} {\bf 384} (2021), no.~1
  279--339, [\href{http://arxiv.org/abs/2001.02177}{{\tt arXiv:2001.02177}}].

\bibitem{Woodhouse:1985id}
N.~M.~J. Woodhouse, {\it {REAL METHODS IN TWISTOR THEORY}},  {\em Class. Quant.
  Grav.} {\bf 2} (1985) 257--291.

\bibitem{Chalmers:1996rq}
G.~Chalmers and W.~Siegel, {\it {The Selfdual sector of QCD amplitudes}},  {\em
  Phys. Rev. D} {\bf 54} (1996) 7628--7633,
  [\href{http://arxiv.org/abs/hep-th/9606061}{{\tt hep-th/9606061}}].

\bibitem{Costello:2021bah}
K.~J. Costello, {\it {Quantizing local holomorphic field theories on twistor
  space}},  \href{http://arxiv.org/abs/2111.08879}{{\tt arXiv:2111.08879}}.

\bibitem{Adamo:2013cra}
T.~Adamo, {\it {Twistor actions for gauge theory and gravity}},  other thesis,
  8, 2013.

\bibitem{Sharma:2021pkl}
A.~Sharma, {\it {Twistor action for general relativity}},
  \href{http://arxiv.org/abs/2104.07031}{{\tt arXiv:2104.07031}}.

\bibitem{Adamo:2013tja}
T.~Adamo and L.~Mason, {\it {Conformal and Einstein gravity from twistor
  actions}},  {\em Class. Quant. Grav.} {\bf 31} (2014), no.~4 045014,
  [\href{http://arxiv.org/abs/1307.5043}{{\tt arXiv:1307.5043}}].

\bibitem{Aganagic:2017tvx}
M.~Aganagic, K.~Costello, J.~McNamara, and C.~Vafa, {\it {Topological
  Chern-Simons/Matter Theories}},  \href{http://arxiv.org/abs/1706.09977}{{\tt
  arXiv:1706.09977}}.

\bibitem{Nair:2005iv}
V.~P. Nair, {\it {A Note on MHV amplitudes for gravitons}},  {\em Phys. Rev. D}
  {\bf 71} (2005) 121701, [\href{http://arxiv.org/abs/hep-th/0501143}{{\tt
  hep-th/0501143}}].

\bibitem{Mason:2010yk}
L.~J. Mason and D.~Skinner, {\it {The Complete Planar S-matrix of N=4 SYM as a
  Wilson Loop in Twistor Space}},  {\em JHEP} {\bf 12} (2010) 018,
  [\href{http://arxiv.org/abs/1009.2225}{{\tt arXiv:1009.2225}}].

\bibitem{Bullimore:2011ni}
M.~Bullimore and D.~Skinner, {\it {Holomorphic Linking, Loop Equations and
  Scattering Amplitudes in Twistor Space}},
  \href{http://arxiv.org/abs/1101.1329}{{\tt arXiv:1101.1329}}.

\bibitem{Chicherin:2014uca}
D.~Chicherin, R.~Doobary, B.~Eden, P.~Heslop, G.~P. Korchemsky, L.~Mason, and
  E.~Sokatchev, {\it {Correlation functions of the chiral stress-tensor
  multiplet in $ \mathcal{N}=4 $ SYM}},  {\em JHEP} {\bf 06} (2015) 198,
  [\href{http://arxiv.org/abs/1412.8718}{{\tt arXiv:1412.8718}}].

\bibitem{Chicherin:2016soh}
D.~Chicherin and E.~Sokatchev, {\it {Demystifying the twistor construction of
  composite operators in ${\mathcal N}=4$ super-Yang\textendash{}Mills
  theory}},  {\em J. Phys. A} {\bf 50} (2017), no.~20 205402,
  [\href{http://arxiv.org/abs/1603.08478}{{\tt arXiv:1603.08478}}].

\bibitem{Adamo:2011cd}
T.~Adamo, {\it {Correlation functions, null polygonal Wilson loops, and local
  operators}},  {\em JHEP} {\bf 12} (2011) 006,
  [\href{http://arxiv.org/abs/1110.3925}{{\tt arXiv:1110.3925}}].

\bibitem{Adamo:2011dq}
T.~Adamo, M.~Bullimore, L.~Mason, and D.~Skinner, {\it {A Proof of the
  Supersymmetric Correlation Function / Wilson Loop Correspondence}},  {\em
  JHEP} {\bf 08} (2011) 076, [\href{http://arxiv.org/abs/1103.4119}{{\tt
  arXiv:1103.4119}}].

\bibitem{Koster:2016fna}
L.~Koster, V.~Mitev, M.~Staudacher, and M.~Wilhelm, {\it {On Form Factors and
  Correlation Functions in Twistor Space}},  {\em JHEP} {\bf 03} (2017) 131,
  [\href{http://arxiv.org/abs/1611.08599}{{\tt arXiv:1611.08599}}].

\bibitem{Koster:2016ebi}
L.~Koster, V.~Mitev, M.~Staudacher, and M.~Wilhelm, {\it {Composite Operators
  in the Twistor Formulation of N=4 Supersymmetric Yang-Mills Theory}},  {\em
  Phys. Rev. Lett.} {\bf 117} (2016), no.~1 011601,
  [\href{http://arxiv.org/abs/1603.04471}{{\tt arXiv:1603.04471}}].

\bibitem{Koster:2016loo}
L.~Koster, V.~Mitev, M.~Staudacher, and M.~Wilhelm, {\it {All tree-level MHV
  form factors in $ \mathcal{N} $ = 4 SYM from twistor space}},  {\em JHEP}
  {\bf 06} (2016) 162, [\href{http://arxiv.org/abs/1604.00012}{{\tt
  arXiv:1604.00012}}].

\bibitem{Cachazo_2004}
F.~Cachazo, P.~Svrcek, and E.~Witten, {\it {MHV vertices and tree amplitudes in
  gauge theory}},  {\em JHEP} {\bf 09} (2004) 006,
  [\href{http://arxiv.org/abs/hep-th/0403047}{{\tt hep-th/0403047}}].

\bibitem{Penrose:1986ca}
R.~Penrose and W.~Rindler, {\em {SPINORS AND SPACE-TIME. VOL. 2: SPINOR AND
  TWISTOR METHODS IN SPACE-TIME GEOMETRY}}.
\newblock Cambridge Monographs on Mathematical Physics. Cambridge University
  Press, 4, 1988.

\bibitem{Costello:2021kiv}
K.~Costello and B.~R. Williams, {\it {Twisted heterotic/type I duality}},
  \href{http://arxiv.org/abs/2110.14616}{{\tt arXiv:2110.14616}}.

\bibitem{Bern:1993qk}
Z.~Bern, G.~Chalmers, L.~J. Dixon, and D.~A. Kosower, {\it {One loop N gluon
  amplitudes with maximal helicity violation via collinear limits}},  {\em
  Phys. Rev. Lett.} {\bf 72} (1994) 2134--2137,
  [\href{http://arxiv.org/abs/hep-ph/9312333}{{\tt hep-ph/9312333}}].

\bibitem{Mahlon:1993si}
G.~Mahlon, {\it {Multi - gluon helicity amplitudes involving a quark loop}},
  {\em Phys. Rev. D} {\bf 49} (1994) 4438--4453,
  [\href{http://arxiv.org/abs/hep-ph/9312276}{{\tt hep-ph/9312276}}].

\bibitem{Cachazo:2004zb}
F.~Cachazo, P.~Svrcek, and E.~Witten, {\it {Twistor space structure of one-loop
  amplitudes in gauge theory}},  {\em JHEP} {\bf 10} (2004) 074,
  [\href{http://arxiv.org/abs/hep-th/0406177}{{\tt hep-th/0406177}}].

\bibitem{Boels:2007gv}
R.~Boels, {\it {A Quantization of twistor Yang-Mills theory through the
  background field method}},  {\em Phys. Rev. D} {\bf 76} (2007) 105027,
  [\href{http://arxiv.org/abs/hep-th/0703080}{{\tt hep-th/0703080}}].

\bibitem{Adamo:2020yzi}
T.~Adamo, L.~Mason, and A.~Sharma, {\it {Gluon scattering on self-dual
  radiative gauge fields}},  \href{http://arxiv.org/abs/2010.14996}{{\tt
  arXiv:2010.14996}}.

\bibitem{Adamo:2022mev}
T.~Adamo, L.~Mason, and A.~Sharma, {\it {Graviton scattering in self-dual
  radiative space-times}},  \href{http://arxiv.org/abs/2203.02238}{{\tt
  arXiv:2203.02238}}.

\end{thebibliography}\endgroup
\end{document}